\documentclass[12pt]{article}

\usepackage{graphicx}%
\usepackage{dcolumn} %
\usepackage{bm}      %
\usepackage{subfigure}
\usepackage{amsfonts}
\usepackage[usenames]{color}
\usepackage{rotating}
\usepackage{fontenc}
\usepackage{amsthm}
\usepackage{hyperref}
\usepackage{cancel}
\usepackage{ulem}

\definecolor{darkred}{RGB}{139,0,0}
\definecolor{chartreuse}{RGB}{127,255,0}
\definecolor{goldenrod}{RGB}{218,165,32}
\definecolor{gray}{RGB}{127,127,127}

\definecolor{Magenta}{RGB}{255, 0,255}
\definecolor{Orange}{RGB}{255,165, 0}
\definecolor{Gray}{RGB}{127,127,127}

\newcommand{\be}{\begin{equation}}
\newcommand{\ee}{\end{equation}}
\newcommand{\bea}{\begin{eqnarray}}
\newcommand{\eea}{\end{eqnarray}}
\newcommand{\bw}{\begin{widetext}}
\newcommand{\ew}{\end{widetext}}
\newcommand{\mm}{\mathrm}

\newcommand{\bi}{\begin{itemize}}
\newcommand{\ei}{\end{itemize}}

\newcommand{\kmean}{\langle k \rangle}

\usepackage{scicite}

\usepackage{times}

\newenvironment{sciabstract}{%
\begin{quote} \bf}
{\end{quote}}

\newcounter{lastnote}
\newenvironment{scilastnote}{%
\setcounter{lastnote}{\value{enumiv}}%
\addtocounter{lastnote}{+1}%
\begin{list}%
{\arabic{lastnote}.}
{\setlength{\leftmargin}{.22in}}
{\setlength{\labelsep}{.5em}}}
{\end{list}}

\title{Revealing the intricate effect of collaboration on innovation}

\author
{Hiroyasu Inoue,$^{1,2}$ Yang-Yu Liu,$^{2,3\ast}$\\
\\
\normalsize{$^{1}$Osaka Sangyo University, Daito, Osaka 574-0013 Japan}\\
\normalsize{$^{2}$Center for Complex Network Research and Department of
  Physics}\\
\normalsize{Northeastern University, Boston, Massachusetts 02115, USA}
\\
\normalsize{$^{3}$Center for Cancer Systems Biology}
\\
\normalsize{Dana-Farber Cancer Institute, Boston, Massachusetts 02115, USA} 
\\
}

\date{}

\begin{document}

\baselineskip24pt

\maketitle

\begin{sciabstract}
We study the Japan and U.S. patent records of several decades to demonstrate
the effect of collaboration on innovation.
We find that statistically inventor teams slightly outperform solo
inventors while company teams perform equally well as solo companies. 
By tracking the performance record of individual teams we find that
inventor teams' performance generally degrades with more repeat 
collaborations. 
Though company teams' performance displays strongly bursty
behavior, long-term collaboration does not significantly help
innovation at all. 
To systematically study the effect of repeat collaboration, we define the repeat collaboration number of a
team as the average number of collaborations over all the
teammate pairs. 
We find that mild repeat collaboration improves the performance of
Japanese inventor teams and U.S. company teams. Yet, excessive repeat
collaboration does not significantly help innovation at
both the inventor and company levels in both countries. 
To control for unobserved heterogeneity, we perform a detailed regression
analysis and the results are consistent with our simple observations. 
The presented results reveal the intricate effect of collaboration on
innovation, which may also be observed in other creative projects. 
\end{sciabstract}

\section*{Introduction}
Collaboration is key to innovation. 
Indeed, collaboration increases the chances of combinations among
ideas, which may result in an innovative and gifted
product~\cite{Simonton-Book-88}. For example, an inventor might combine
his or her half idea with another inventor's half idea to realize a
whole innovative one. 
Moreover, collaboration can speed up the delivery of
innovations~\cite{Weisberg-Book-06}, which may involve the parallel
validation of initial conceptions and the series implementation of
final ideas. Since speed is the last great competitive advantage to
innovations, the speed-up gained through collaboration could be a crucial
determinant in creative enterprises. 
While collaboration has been considered as a central theme to
innovation, the effect of collaboration on innovation has not been
quantitatively studied in a systematic fashion. %

Previous studies found that
repeat collaborations usually underperform in creative projects,
e.g., scientific research~\cite{Guimera05,Porac-RP-04},
consulting practice~\cite{Reagans-ASQ-04}, and entertainment
performances~\cite{Guimera05,Delmestri-JMS-05,Perretti-JOB-07,Uzzi-AJS-05}. 
Those interesting results were explained by the suppression of ``creative
abrasion'' (a sequence of processes constituted by idea generation,
disclosure/advocacy, and convergence), which is key to creative
project performance~\cite{Skilton-AMR-10}. 
Despite those intriguing results on the negative relationship between repeat
collaboration and team performance, the effect of repeat collaboration on
innovation has not been fully understood. %

Here we study the Japan and U.S. patent records of several decades~\cite{japio,nber,iqss,Hall00,Tamada02} to
demonstrate the effect of collaboration on innovation.  
Patent records are %
valuable for this research. 
First of all, the purpose of patents is to facilitate and encourage disclosure of
innovations into the public domain for the common good. 
A typical patent application must meet the relevant patentability
requirements such as novelty and non-obviousness. 
Hence, patent records are directly related to the
occurrence of innovations over time~\cite{Griliches98,Fleming-RP-01,Fleming-ASQ-07}.
We can track and analyze the innovation activity over long periods of time by mining
patent records, in line with the current quantitative trend of computational social
science~\cite{Lazer-Science-09}. 
Comparing with patent records, team performance in scientific
research, consulting practice, and entertainment industries, cannot
always be directly related to innovation. 
For example, %
scientific findings, especially from fundamental sciences, do not always lead
to more effective products or technologies that are readily available to
markets and society. 
Second, there are two levels of collaboration in patent
records. A patent application can be filed by multiple inventors or/and
multiple companies. 
Though different companies could have different climates and unique tacit
knowledge~\cite{Nonaka94}, to speed up innovations companies capitalize on
other companies' knowledge more and
more~\cite{Chesbrough03,Laursen06,Grant96}. 
Commensurate with this trend, the number of joint patents applied by
multiple companies keeps increasing those days~\cite{Hicks00}. 
Since innovations can be driven by the collaborations of inventors or/and
companies, it would be very interesting to study the effect of collaboration
on innovation at both the inventor and company levels. 
Patent records can hence help us understand the difference and/or
similarity of collaborationship at different organization levels.

\section*{Collaboration networks} 
A patent can be requested by filing an application. The applicant of a patent
may be inventors or/and companies. 
In this work we analyze the Japan and U.S. patent records, which cover different years and different number of inventors and
companies. We first study the structure of the underlying
collaboration networks to check the similarity and/or difference of the
two patent records. %
We construct a collaboration network of inventors (or companies) by
drawing a link between two nodes $i$ and $j$ if they collaborate at
least once, i.e., they file at least one patent application together
(see Fig.1), where nodes are inventors (or companies) and links
represent the collaborations between inventors (or
companies)~\cite{Fleming-CMR-06}. 
The total number of collaborators of node $i$ is called its degree,
denoted as $k_i$. 
The total times of collaborations between
nodes $i$ and $j$ is defined to be the weight of the link $(i,j)$,
denoted as $w_\mm{l}(i,j)$. %
The total number of patents that node $i$ has contributed is defined to be its
weight, denoted as $w_\mm{n}(i)$. 
Table.\ref{table:CN} shows the basic information of the Japan and U.S. patent
records and the constructed collaboration networks. 
We find that at the inventor level both Japan and U.S. collaboration
networks show very high clustering coefficient $C$ and high
assortative degree correlations $r$.  High $C$ indicates that inventors tend
to cluster together, i.e., an inventor's two collaborators also tend to be
collaborators of each other. High $r$ means that hub inventors (with high
degree $k$) tend to collaborate with other hub inventors. 
At the company level, however, both Japan and U.S. collaboration networks
display very low clustering coefficient and slightly disassortative degree
correlations, which are qualitatively different from the inventor
collaboration networks. 
Despite the fact that Japan and U.S. collaboration networks cover
different years and different number of inventors and companies, we
find that their degree distributions $P(k)$, node weight distributions
$P(w_\mm{n})$, link weight distributions $P(w_\mm{l})$, and component size
distributions $P(S)$ display qualitatively similar
features (see SOM for details). 
At the company level, we do find the Japan and U.S. collaboration
networks display quantitative differences. For example, they have different 
fractions of isolated nodes ($n_0=0.542$ for Japan and $0.907$ for U.S.) who
never collaborate with others.  
Their largest connected component sizes are also different
($s_\mm{lc}=0.364$ for Japan and $0.049$ for U.S.).  
The presence of the giant component which occupies a finite fraction of nodes
in the Japan company collaboration network indicates that Japanese companies
are highly connected through collaboration. In contrast the U.S. companies are
still not well connected in innovative teams. 
This structure difference is also reflected by their mean degrees
  ($\kmean=1.941$ for Japan and $0.214$ for U.S.).
The high value of $n_0$, low values of $s_\mm{lc}$ and $\kmean$ for
the U.S. company collaboration network implies that company collaborations in
innovations are not very popular in U.S. 
Note that according to both the U.S. patent laws (35 U.S.C. 262) and Japanese
patent law (Article 73), a company cannot sell or license a jointly applied
patent without the consent of others. 
Hence, the joint application of patents would usually be considered as
a second-best option~\cite{Hagedoorn03}. 
Yet, Japan companies seems to be more open to collaborate on patents than
U.S. companies. %
\section*{Collaboration and innovation}

\subsection*{Effect of Team Size}
We first illustrate the effect of team size on innovation. Previous
studies have shown that inventor teams typically produce more
successful patents than solo inventors do
~\cite{Wuchty07,Fleming10}. Yet, it is still unknown whether company
teams will also outperform solo companies. 
We denote the number of inventors (or companies) listed in a patent record as
$m$. An inventor (or company) team is defined as having more than one listed
inventor (or company) in a patent record (i.e., $m \ge 2$). 
To quantify the innovation performance of inventors and companies, we
define the impact ($I$) of a patent to be the number of
citations of that patent normalized by the average number of citations
of patents granted in the same year~\cite{Trajtenberg90,Hall00}. 
We find that in average inventor teams outperform solo inventors (see Fig.\ref{fig:team}a),
consistent with previous result~\cite{Wuchty07}. However, at the
company level in average teams does not outperform solos at all (see
Fig.\ref{fig:team}f). 
In fact, the average patent impact of the U.S. company teams is even less than
that of U.S. solo companies.  
To further compare the performance of solos and teams, we calculate the impact
distributions $P(I)$ of patents invented by solos and teams, separately. We
find that $P(I)$ displays fat-tailed distributions at both the inventor and
company levels, consistent with the result of $P(I)$ calculated for all patent
records regardless of whether they are filed by inventors or companies
(see SOM).  
In particular, we observe that at the inventor level teams are more likely to have patents with huge impact than solos; while at the company level teams do not show such outstanding performance comparing to solos.

To reveal more information about the effect of team size on innovation, we
systematically study the patent impact ($I$) as a function of team 
size ($m$). 
We find that the team performance, as measured by the impact of their
patents,  behaves differently at the two different levels as the team size increases. 
For inventor teams, the patent impact increases slowly as team size $m$
increases (up to $m \sim 15$), especially for the Japanese inventor teams,
consistent with the results shown in Fig.\ref{fig:team}a.  
For company teams, however, the patent impact does not increase significantly
with increasing $m$, consistent with the results shown in
Fig.\ref{fig:team}f. 
We also notice that for both inventor and company teams their
performance displays large fluctuations with large team size $m$, which could be
due to the fact that large teams are rather rare in both Japan and U.S. patent
records (see Fig.\ref{fig:team}d,i). 

\subsection*{Effect of Team Experience}
\subsubsection*{Repeat Collaboration}

Team experience is another important factor that could potentially
affect a team's innovation performance. %
To demonstrate the effect of team experience on innovation, one
can simply track the performance of each team. %
For a given team, represented by a set of inventors (or companies), we 
define its \emph{exact repetition number} ($R$) %
as the accumulated number of patent applications that the whole team
has filed together up to the current patent. 
We then label teams %
according to their inventor (or company) set and track each team's
performance by plotting the impact of their patents as a function of 
$R$  (see Fig.~\ref{fig:trackrecord}). %
Interestingly, we find that extremely successful patents (indicated by
their huge impact) are typically among the first 10 patents of most
inventor teams, i.e., $R\le 10$ .
For company teams, their patent records display many impact spikes or bursts,
indicating that individual company teams occasionally perform extremely
well. (Note that this strongly bursty behavior is not noticeable at the inventor
level.) Yet, for both Japan and U.S. company teams long-term
collaboration does not significantly help innovation at all. In fact for Japan 
company teams we see a trend that the performance degrades as 
$R$ increases.

\begin{figure*}[t!]
\begin{center}
\includegraphics[width=\textwidth]{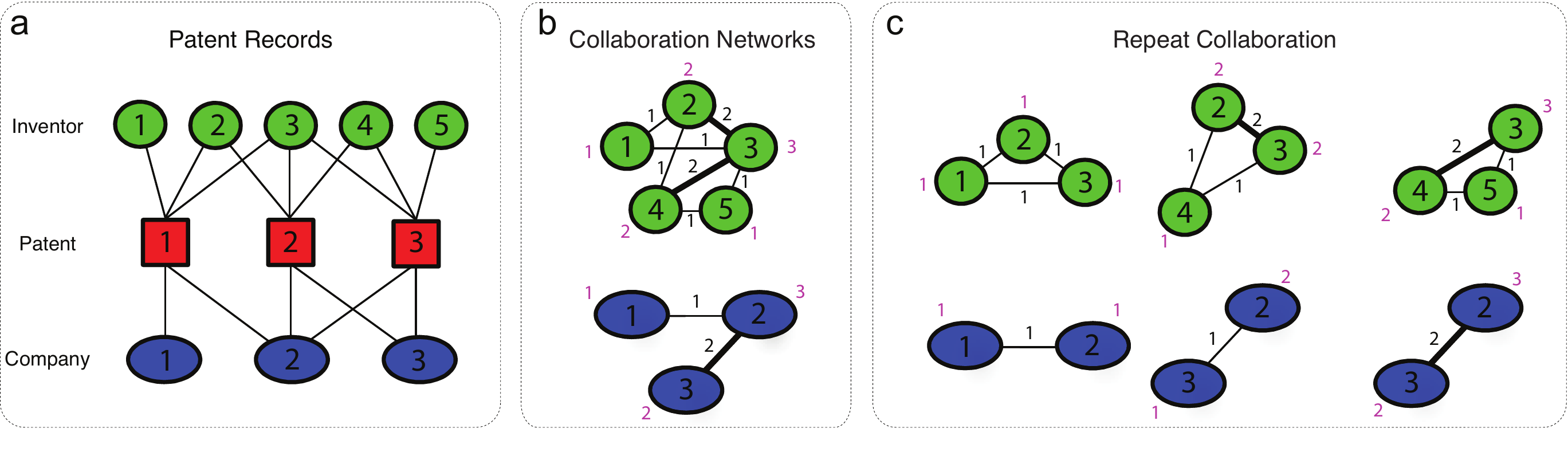}
\caption{{\bf Patent records and the associated collaboration
    networks.} 
(a) Patent records contain collaborations at both the inventor and company
  levels.  
(b) By drawing a undirected link between nodes $i$ and
$j$ if they file a patent application at least once, we can
construct the collaboration networks of inventors (or companies). 
The total times of collaborations between nodes $i$ and $j$ over the whole
patent record is defined to be the weight of the link $(i,j)$ (shown in black). 
The total number of collaborators of nodes $i$ over the whole patent record is
defined to be the weight of the node $i$ (shown in pink). 
(c) The inventors (or companies) listed in each patent record forms a clique.  
For each patent, we calculate its repeat collaboration number
($R_\mm{l}$) of its inventors (or companies) by averaging the accumulated number of collaborations among of all the inventor (or
company) pairs in the team (shown in black). 
The productivity of node $i$ in a patent is defined to be the accumulated number of patents that
node $i$ has contributed. 
We calculate the team productivity ($R_\mm{n}$) by averaging the
productivity of all its nodes (shown in
pink). 
} \label{fig:schema}
\end{center}
\end{figure*}

\begin{figure*}[t!]
\begin{center}
\begin{minipage}[t]{0.48\textwidth}
\centering
\includegraphics[width=\textwidth,height=0.56\textwidth]{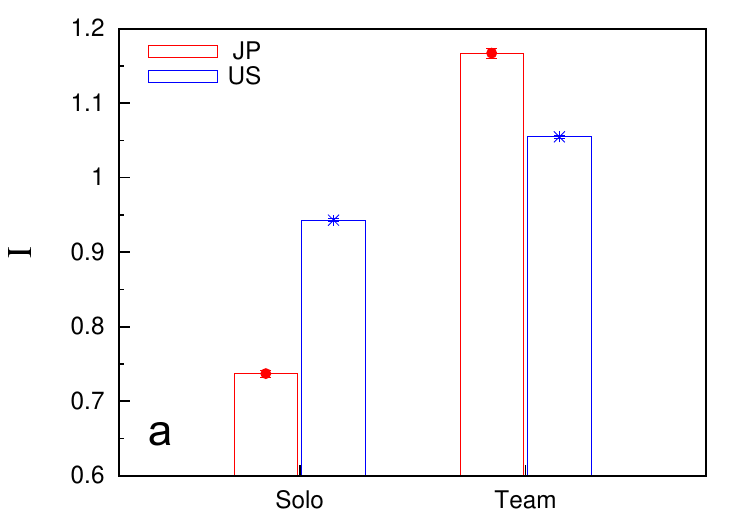}
\includegraphics[width=0.478\textwidth,height=0.56\textwidth]{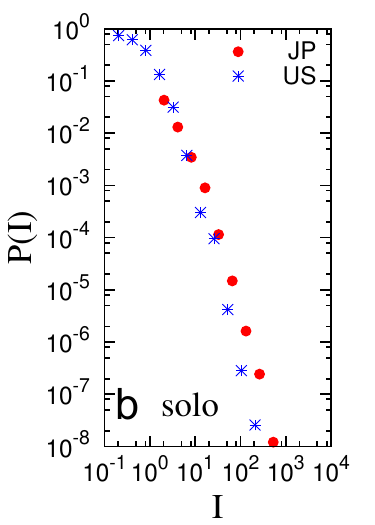}
\includegraphics[width=0.478\textwidth,height=0.56\textwidth]{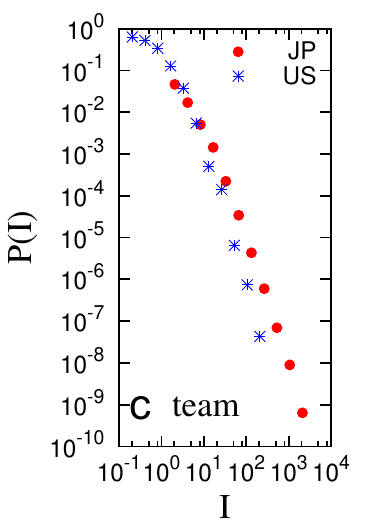}
\end{minipage}
\begin{minipage}[t]{0.48\textwidth}
\centering
\includegraphics[width=\textwidth,height=0.56\textwidth]{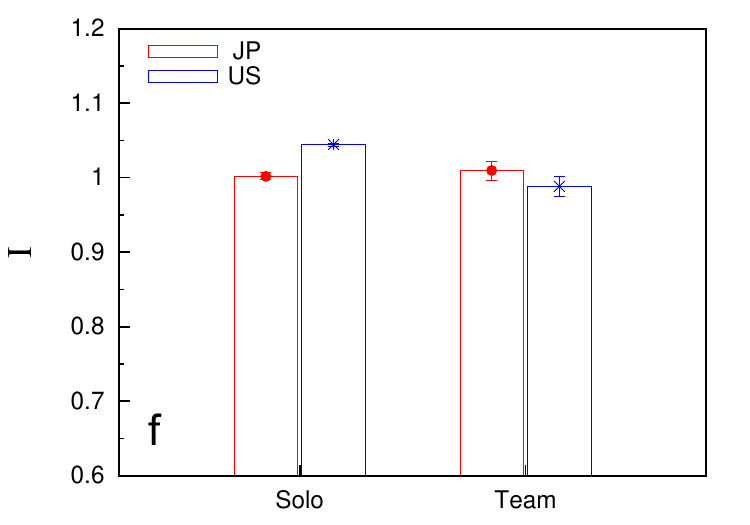}
\includegraphics[width=0.478\textwidth,height=0.56\textwidth]{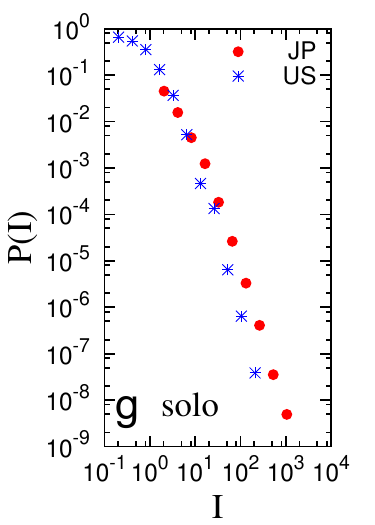}
\includegraphics[width=0.478\textwidth,height=0.56\textwidth]{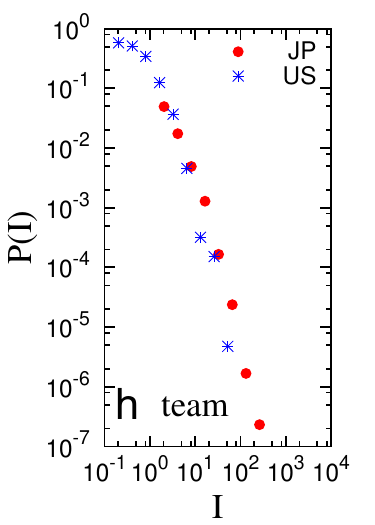}
\end{minipage}\\
\vspace{0.25in}
\begin{minipage}[t]{0.48\textwidth}
\includegraphics[width=\textwidth,height=0.6\textwidth]{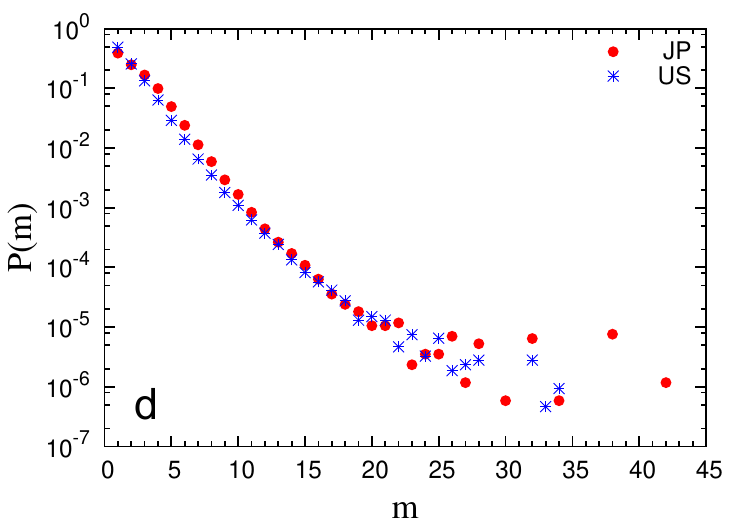}
\includegraphics[width=\textwidth,height=0.6\textwidth]{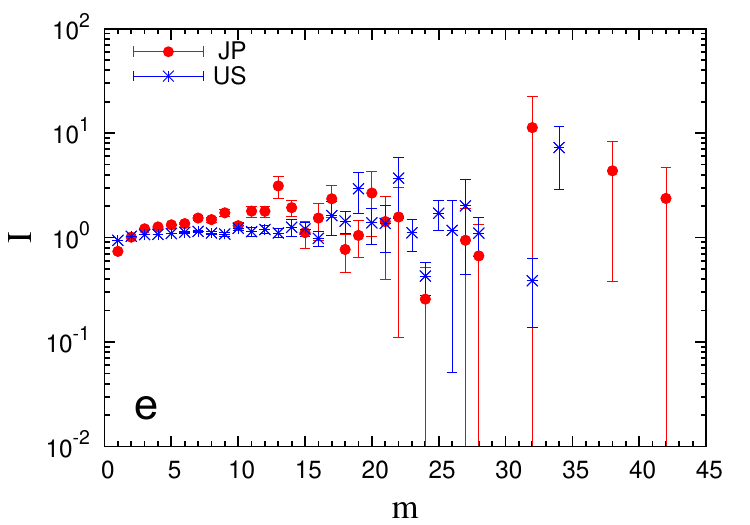}
\end{minipage}
\begin{minipage}[t]{0.48\textwidth}
\includegraphics[width=\textwidth,height=0.6\textwidth]{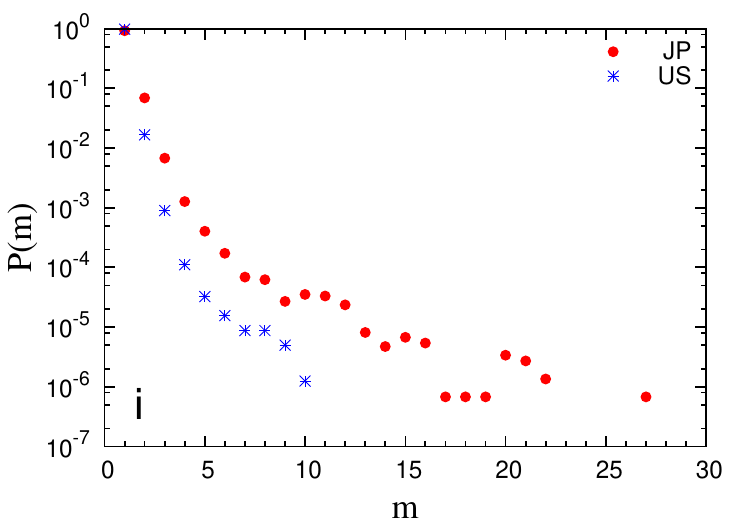}
\includegraphics[width=\textwidth,height=0.6\textwidth]{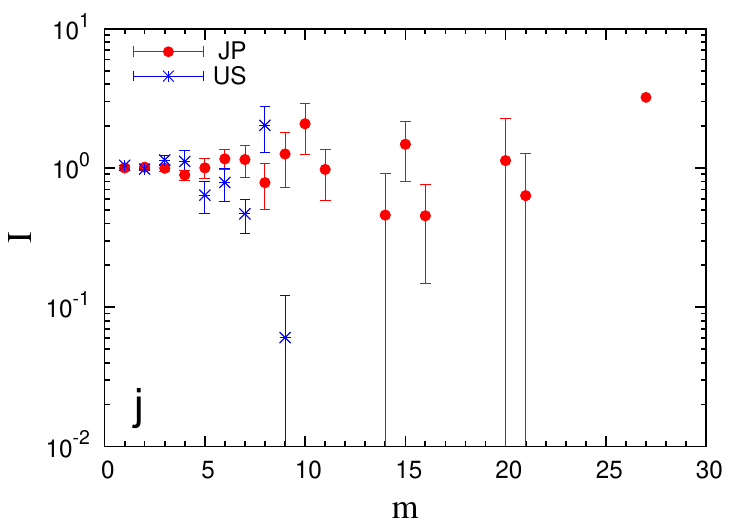}
\end{minipage}
\end{center}
\caption{{\bf Effect of team size on innovation performance.} 
  (a-e) Inventors. (f-j) Companies. 
(a,f) The average impacts of patents filed by solos and teams.
(b,c,g,h) 
The impact distribution of patents filed by solos ($m=1$)
  and teams ($m\ge 2$).
(d,i) The team size distribution.
(e,j) The patent impact as a function of team size. 
}\label{fig:team}
\end{figure*}

\begin{figure*}[!t]
\begin{center}
\includegraphics[width=0.49\textwidth,height=0.3\textwidth]{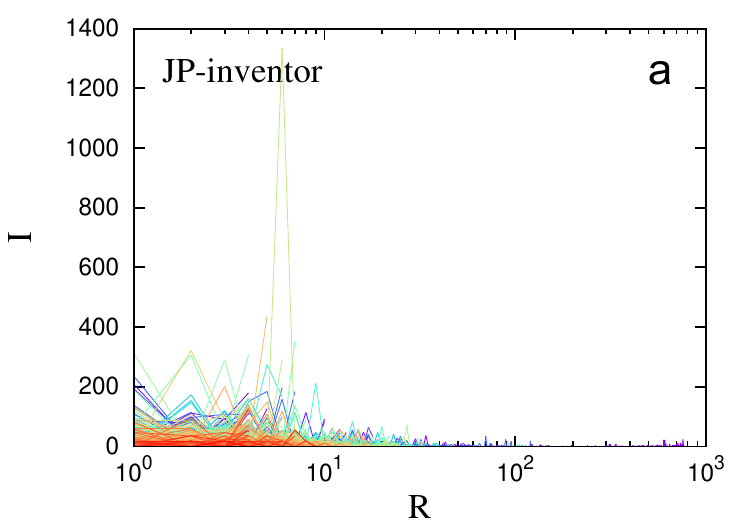}
\includegraphics[width=0.49\textwidth,height=0.3\textwidth]{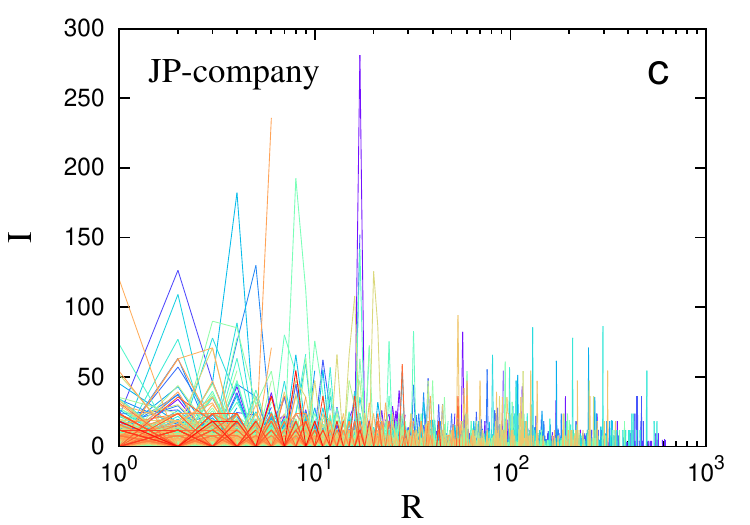}
\includegraphics[width=0.49\textwidth,height=0.3\textwidth]{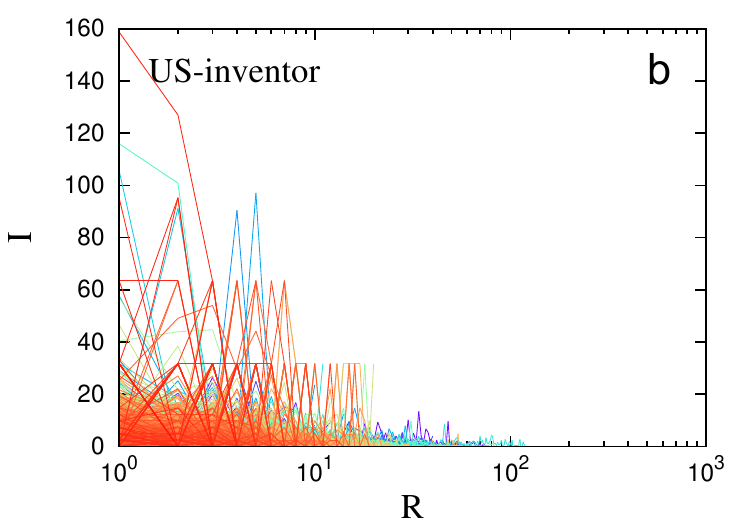}
\includegraphics[width=0.49\textwidth,height=0.3\textwidth]{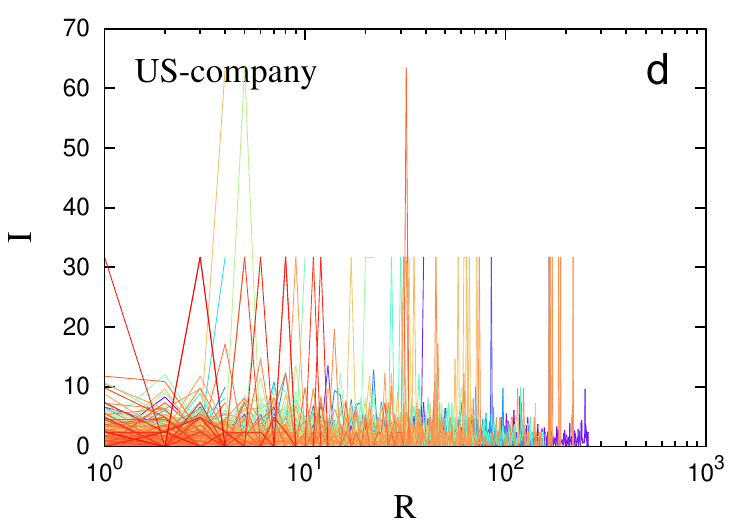}
\end{center}
\caption{{\bf Track records of individual teams with at least three patent
    records.} Different colors represent different individual teams. 
  (a,b) Inventors. (c,d) Companies. 
}\label{fig:trackrecord}  
\end{figure*}

In the above analysis we focus on the repeat collaboration
of the whole team rather than its members. Hence only a small
portion of patent records is analyzed.  
Actually before the whole team work together again, some
of its team members may have already collaborated or worked alone on some
other patents.  
To take this into account and systematically study the effect of
repeat collaboration on innovation using all the patent records of teams, we
denote $R_{ij}$ of a node pair ($i,j$) in a patent record as the 
accumulated number of repeat collaborations between $i$ and $j$ up to
that patent. 
We then define the \emph{repeat collaboration number} ($R_\mm{l}$) of a
team listed in a patent record as the average $R_{ij}$ of all its
teammate pairs.  
For example, in Fig.\ref{fig:schema} 
the repeat  collaboration number of the inventor team in patent-2 is
$R_\mm{l} = (1+1+2)/3=4/3$. 

\begin{figure*}[!t]
\begin{center}
\begin{minipage}[b]{0.49\textwidth}
\begin{flushright}
\includegraphics[width=0.978\textwidth,height=0.6\textwidth]{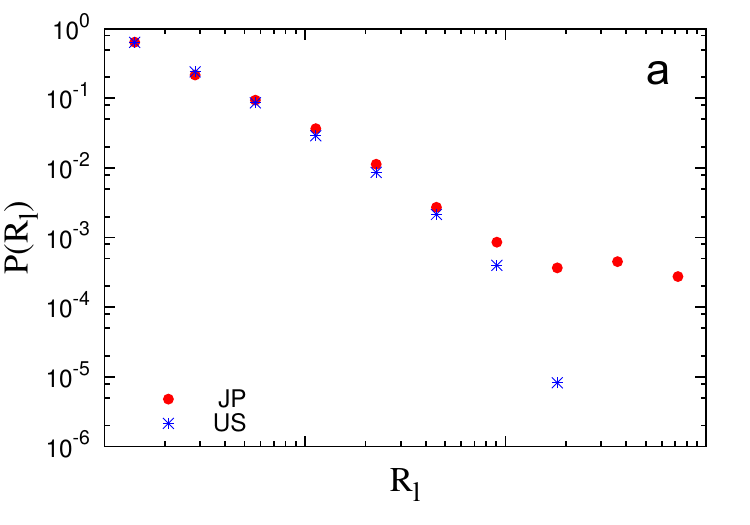}
\includegraphics[width=\textwidth,height=0.6\textwidth]{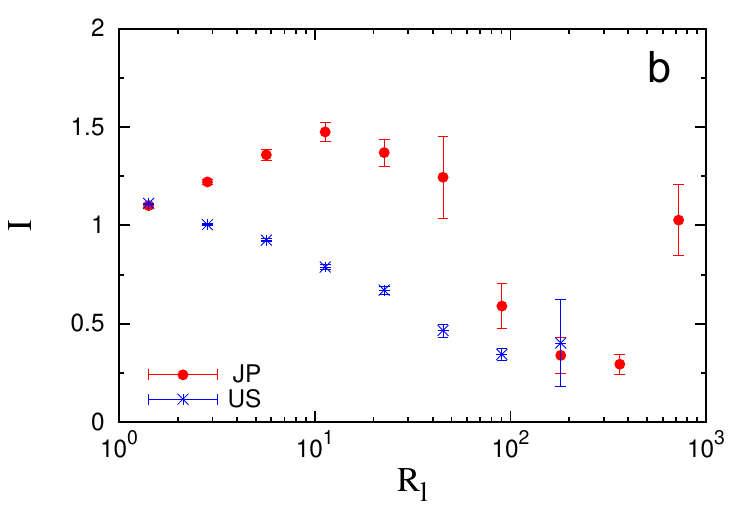}
\end{flushright}
\end{minipage}
\begin{minipage}[b]{0.49\textwidth}
\begin{flushright}
\includegraphics[width=0.978\textwidth,height=0.6\textwidth]{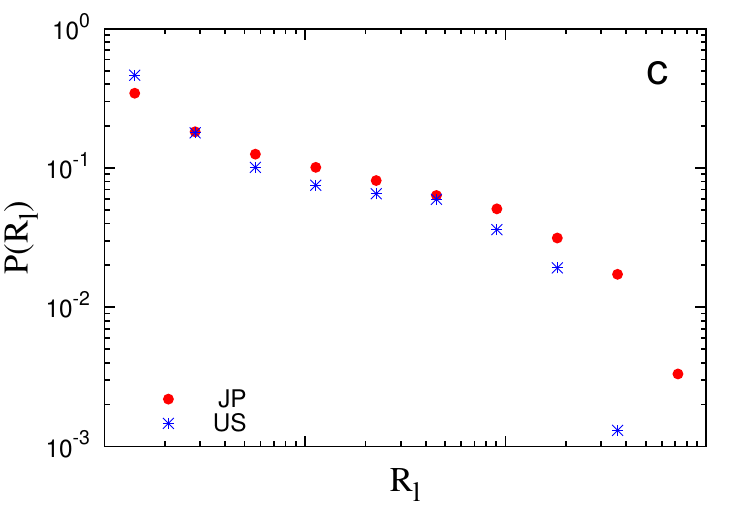}
\includegraphics[width=\textwidth,height=0.6\textwidth]{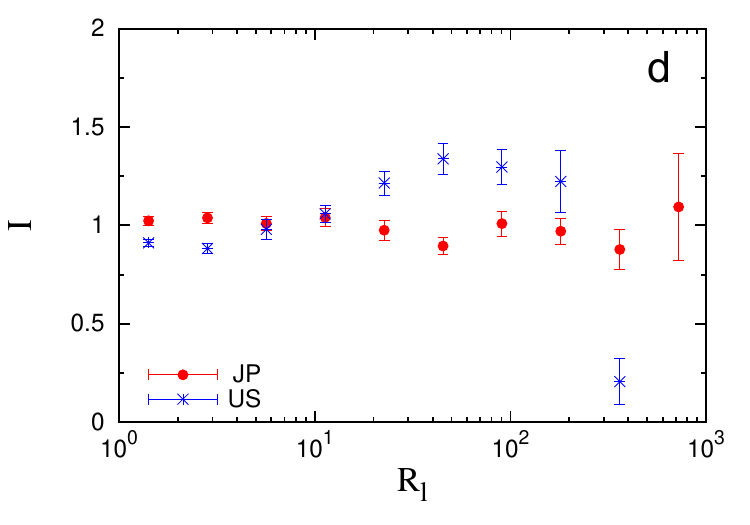}
\end{flushright}
\end{minipage}
\end{center}
\caption{{\bf Effect of repeat collaboration on innovation performance.} 
(a,b) Inventors. (c,d) Companies. 
(a,c) Distribution of repeat collaboration number of teams. 
  (b,d) Patent Impact as a function of repeat collaboration number.
}\label{fig:impact-age}  
\end{figure*}

For each patent in the patent records of teams, we calculate its $R_\mm{l}$ 
and find that $R_\mm{l}$ shows a broad distribution for both
inventor and company teams (see Fig.\ref{fig:impact-age}a,c). 
We then calculate the average patent impact for teams of similar
$R_\mm{l}$ grouped in logarithmic bins (see Fig.\ref{fig:impact-age}b,d). 
Interestingly, we find that the effects of repeat collaboration at the
inventor and company levels are qualitatively different.  
At the inventor level we find Japanese teams and U.S. teams also display
quite different behavior. 
The innovation performance of Japanese inventor teams improves first as
$R_\mm{l}$ increases, reaches its peak value at $R_\mm{l}=10$, and then
generally degrades for $R_\mm{l} > 10$ (except the abnormal behavior around
$R_\mm{l}\approx 700$, where the patent impacts increases but is still not
significantly higher than that of teams with $R_\mm{l}<10$.) 
This suggests an ideal timing for Japanese inventors to make new
collaborations and hence ``rejuvenate'' the inventor team. 
In contrast, the performance of U.S. inventor teams degrades almost
monotonically as $R_\mm{l}$ increases, implying that repeat
collaborations weakens the creativity of U.S. inventor teams.  
At the company level Japanese teams show remarkably stable performance
for $R_\mm{l}$ up to $10^3$. 
For U.S. company teams their performance slightly improves as $R_\mm{l}$
increases up to $100$ and then degrades with increasing $R_\mm{l}$. 
Neither Japanese nor U.S. company teams perform significantly well with long
term collaborations. 

\subsubsection*{Regression Analysis}
Besides the exact repetition number ($R$) and the repeat
collaboration number ($R_\mm{l}$) of a team, there are numerous
variables related to team experience, e.g., \emph{team age} (denoted
as $A$, the
average of the team members' ``age'', i.e., the duration from its first
application year to the current application year), \emph{team
  productivity} (denoted as $R_\mm{n}$,
the average number of patents that inventors/companies of a team have
already applied), etc (see SOM Table. SI for all the explanatory variables related
to team experience). To control for unobserved
heterogeneity, we performed a detailed regression
analysis~\cite{Fleming10} to investigate the effects of those team experience variables (see SOM for details).    
Interestingly, by calculating the Akaike information criterion (AIC)
value of the statistical models including different sets of variables,
we find that team age ($A$) and repeat collaboration number
($R_\mm{l}$) are better than other team experience variables in
explaining the data sets.  Moreover, we find that $A$ and $R_\mm{l}$ show significance for
all data sets (Japan-inventor, US-inventor, Japan-company,
and US-company), but other team experience variables do not. 

\subsubsection*{Interplay between team age and repeat collaboration}
\begin{figure*}[!t]
\begin{center}
   \includegraphics[width=0.49\textwidth,height=0.3\textwidth]{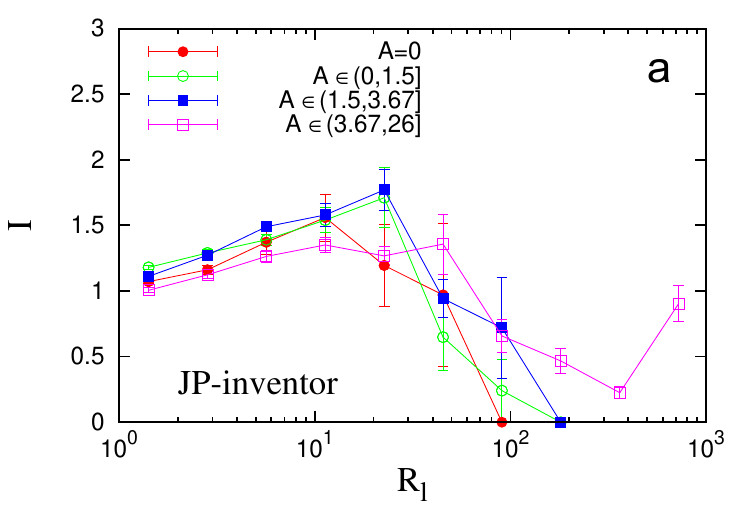}
   \includegraphics[width=0.49\textwidth,height=0.3\textwidth]{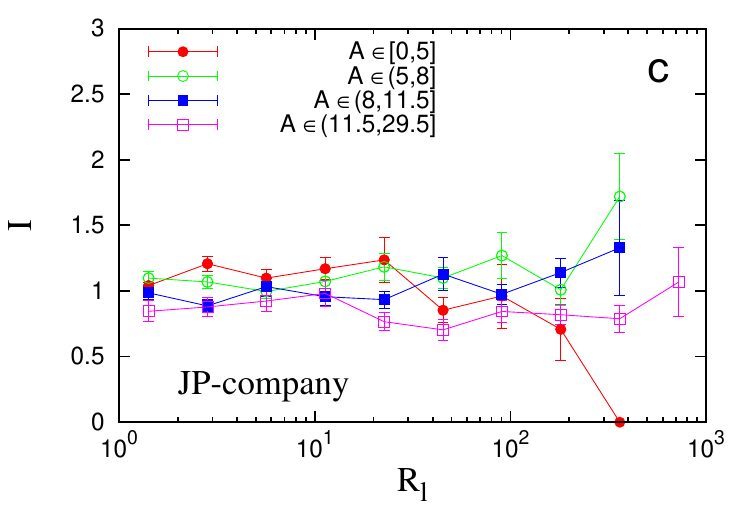}
    \includegraphics[width=0.49\textwidth,height=0.3\textwidth]{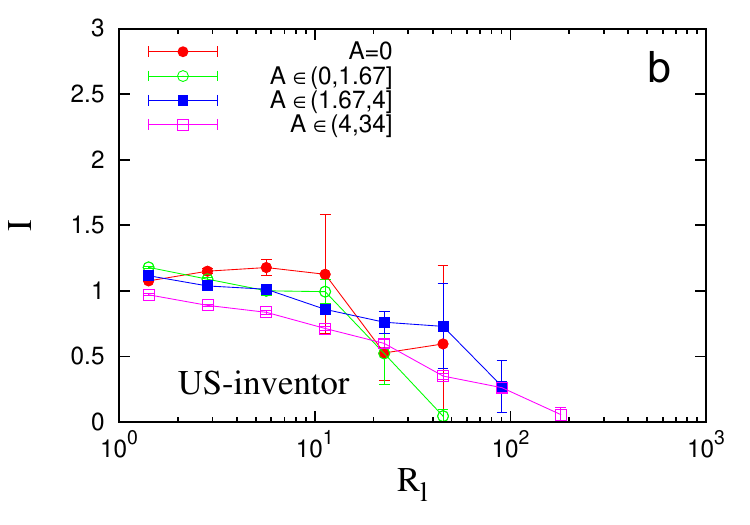}
   \includegraphics[width=0.49\textwidth,height=0.3\textwidth]{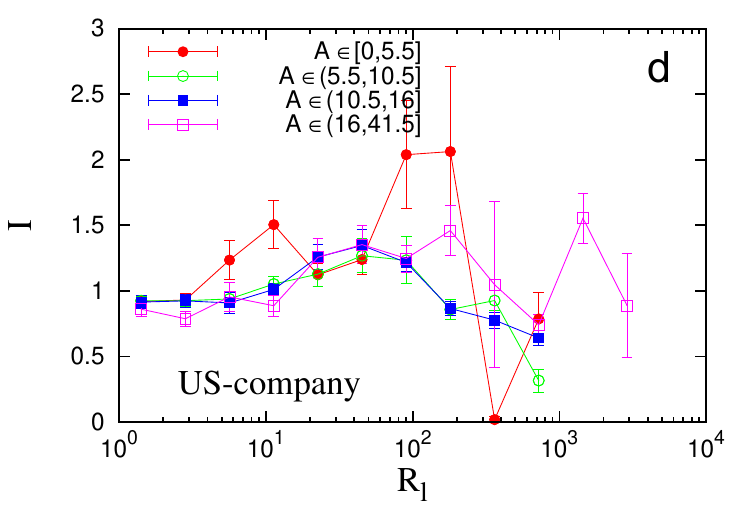}
\end{center}
\caption{{\bf Effect of repeat collaboration on innovation performance
    of teams with similar team age.} 
  (a,b) Inventors. (c,d) Companies. 
  To separate the aging effect of a team from that
  of repeat collaboration among its teammates, we group patents according to
  the quartiles of their team age ($A$). 
  The $A$-range of each group is shown in the legend. 
  For each $A$-group we further group patents according to
  their repeat collaboration number ($R_\mm{l}$) and then calculate
  the average patent impact for each $R_\mm{l}$-subgroup. 
}\label{fig:impact-age-2D}  
\end{figure*}

The result of regression analysis prompts us to study the interplay
between $A$ and $R_\mm{l}$, i.e., the ``aging'' of team members and
the repeat collaboration among them.
Naturally the degradation of team performance with large $R_\mm{l}$
could be possibly due to the fact when teams become older (i.e., $A$
is very large) they are less innovative.   
To address this issue and further reveal the intricate effect of
collaboration on innovation, we group
patents according to quartiles of their team age and then for each
group we calculate the patent impact as a function of
$R_\mm{l}$ (see Fig.\ref{fig:impact-age-2D}). 
Now within each group the change of
innovation performance is mainly due to the repeat collaborations
quantified by $R_\mm{l}$.     
At the inventor level, we find that for Japanese inventor teams,
regardless of their team age, their performance improves first as
$R_\mm{l}$ increases and then degrades. 
Interestingly, U.S. inventor teams' performance degrades almost
monotonically with increasing $R_\mm{l}$, regardless of their
team age. 
At the company level, we find that Japanese company
teams of different team age show remarkably stable innovation
performance as $R_\mm{l}$ increases. In contrast, U.S. company
teams' performance displays quite unstable behavior and there is no
significant improvement for large $R_\mm{l}$, regardless of
the team age. 

\section*{Discussion}
Though for Japanese inventor teams and U.S. company teams a moderate
repeat collaboration slightly improves their innovation performance, 
overall we didn't find strongly positive relationship between
innovation and collaborations in the long run. 
Current results actually suggest that there is a negative relationship between
them, especially at the inventor level and for long term collaboration.  

At the inventor level, we observe that Japanese inventor teams have a performance 
peak around repeat collaboration number $R_\mm{l}\approx 10$ %
while for U.S. inventor teams their innovation performance drops
almost monotonically
as $R_\mm{l}$ increases. 
This result raises an interesting question worthy of future pursuit: 
What causes the drastically different effects of repeat collaboration
on the performance of the two countries' inventor teams? 
We leave the systematic study of this question as future work. 
Here we want to point out the different innovation climates in U.S. and
Japan, which might help us better understand this question. Typically, U.S. workers are subject to the strong
pressure/incentive for the immediate result and light regulations from
the labor market~\cite{Bloom10}, implying that taking time for
U.S. inventors to deepen their collaborations is not a good
strategy. In contrast, the labor regulation of Japan is strict and a
group of individuals can create value among them due to cohesive
culture~\cite{Ralston07}.

At the company level, we observe that Japanese company teams display remarkably
stable innovation performance while U.S. company teams slightly outperform
with repeat collaborations up to some point.  
This might be related to the fact that in U.S. joint
patents of companies are still not very popular, while %
in Japan joint patents of companies have been prevailing. 
Of course, a deeper understanding deserves a systematic study in the
future. Here we emphasize that the difference of intercompany relationships in the two
countries could be useful to explain the observation. In Japan there
is a unique company ties called ``Keiretsu'',
i.e., a set of companies with interlocking business relationships and
shareholdings and hence they typically share human assets and
information~\cite{Dyer96}. Because ``Keiretsu'' significantly
eliminates the difference of inter-company and inner-company, there
may be no margin to deepen their collaborations on innovation. In contrast,
U.S. companies do not have such prior connections and hence they could
build deeper collaborations as they have more joint
patents. Consequently, the longitudinal relationship of U.S. companies
nurture the trust and therefore better
performance~\cite{Zaheer98}. Yet, in the long run, overembeddedness
limits the diversity of information and hence stifles the
creation~\cite{Uzzi97}. This explains the non-monotonic behavior of
the innovation performance of U.S. company teams.

The results presented here provide us a novel perspective about the strategy of improving
innovation performance via controlling the %
repeat collaboration number at both inventor and
company levels. %
Our systematic approach based on team sizes and repeat collaboration can
also be readily applied to other creative projects, such as scientific
research~\cite{Guimera05, Porac-RP-04}, consulting
practice~\cite{Reagans-ASQ-04}, and entertainment
performances~\cite{Delmestri-JMS-05, Guimera05, Perretti-JOB-07,Uzzi-AJS-05},
to further reveal the intricate relation between collaboration and creativity. 

\bibliography{references}

\begin{thebibliography}{10}

\bibitem{Simonton-Book-88}
D.~K. Simonton, {\it Scientific Genius: A Psychology of Science\/} (Cambridge
  University Press, 1988).

\bibitem{Weisberg-Book-06}
R.~W. Weisberg, {\it Creativity: Understanding Innovation in Problem Solving,
  Science, Invention, And the Arts\/} (John Wiley \& Sons Inc, 2006).

\bibitem{Guimera05}
R.~Guimera, B.~Uzzi, J.~Spiro, L.~Amaral, {\it Science\/} {\bf 308}, 697
  (2005).

\bibitem{Porac-RP-04}
J.~F. Porac, {\it et~al.\/}, {\it Research Policy\/} {\bf 33}, 661  (2004).

\bibitem{Reagans-ASQ-04}
R.~Reagans, E.~Zuckerman, B.~McEvily, {\it Administrative Science Quarterly\/}
  {\bf 49}, pp. 101 (2004).

\bibitem{Delmestri-JMS-05}
G.~Delmestri, F.~Montanari, A.~Usai, {\it Journal of Management Studies\/} {\bf
  42}, 975 (2005).

\bibitem{Perretti-JOB-07}
F.~Perretti, G.~Negro, {\it Journal of Organizational Behavior\/} {\bf 586},
  563 (2007).

\bibitem{Uzzi-AJS-05}
B.~Uzzi, J.~Spiro, {\it American Journal of Sociology\/} {\bf 111}, pp. 447
  (2005).

\bibitem{Skilton-AMR-10}
P.~F. Skilton, K.~J. Dooley, {\it Academy of Management Review\/} {\bf 35}, 118
  (2010).

\bibitem{japio}
Http://www.japio.or.jp/english/.

\bibitem{nber}
Http://www.nber.org/.

\bibitem{iqss}
Ronald Lai; Alexander D'Amour; Amy Yu; Ye Sun; Lee Fleming, 2011,
  "Disambiguation and Co-authorship Networks of the U.S. Patent Inventor
  Database (1975 - 2010)", http://hdl.handle.net/1902.1/15705
  UNF:5:RqsI3LsQEYLHkkg5jG/jRg== V3 [Version].

\bibitem{Hall00}
B.~Hall, A.~Jaffe, The nber patent citations data file: Lessons, insights and
  methodological tools, National Bureau of Economic Research Working Paper 8498
  (2000).

\bibitem{Tamada02}
S.~Tamada, F.~Kodama, K.~Gemba, {\it The Journal of Science Policy and Research
  Management\/} {\bf 17}, 222 (2002).

\bibitem{Griliches98}
Z.~Griliches, {\it R\&D and Productivity-The Economic Evidence\/} (The
  University of Chicago Press, Chicago, 1998).

\bibitem{Fleming-RP-01}
L.~Fleming, O.~Sorenson, {\it Research Policy\/} {\bf 30}, 1019 (2001).

\bibitem{Fleming-ASQ-07}
L.~Fleming, S.~Mingo, D.~Chen, {\it Administrative Science Quarterly\/} {\bf
  52}, 443 (2007).

\bibitem{Lazer-Science-09}
D.~Lazer, {\it et~al.\/}, {\it Science\/} {\bf 323}, 721 (2009).

\bibitem{Nonaka94}
I.~Nonaka, A.~Lewin, {\it Organization Science\/} {\bf 5}, 14 (1994).

\bibitem{Chesbrough03}
H.~W. Chesbrough, {\it Open innovation\/} (Harvard Business School, 2003).

\bibitem{Laursen06}
K.~Laursen, A.~Salter, {\it Strategic Management Journal\/} {\bf 27}, 131
  (2006).

\bibitem{Grant96}
R.~Grant, {\it Strategic management journal\/} {\bf 17}, 109 (1996).

\bibitem{Hicks00}
D.~Hicks, A.~{Breitzman Sr}, K.~Hamilton, F.~Narin, {\it Science and Public
  Policy\/} {\bf 27}, 310 (2000).

\bibitem{Fleming-CMR-06}
L.~Fleming, M.~Marx, {\it California Management Review\/} {\bf 48}, 6 (2006).

\bibitem{Hagedoorn03}
J.~Hagedoorn, H.~Kranenburg, R.~Osborn, {\it Managerial and Decision
  Economics\/} {\bf 24}, 71 (2003).

\bibitem{Wuchty07}
S.~Wuchty, B.~Jones, B.~Uzzi, {\it Science\/} {\bf 316}, 1036 (2007).

\bibitem{Fleming10}
J.~Singh, L.~Fleming, {\it Management Science\/} {\bf 56}, 41 (2010).

\bibitem{Trajtenberg90}
M.~Trajtenberg, {\it Rand Journal of Economics\/} {\bf 21}, 172 (1990).

\bibitem{Bloom10}
N.~Bloom, J.~Reenen, {\it The Journal of Economic Perspectives\/} {\bf 24}, 203
  (2010).

\bibitem{Ralston07}
D.~Ralston, D.~Holt, R.~Terpstra, Y.~Kai-Cheng, {\it Journal of International
  Business Studies\/} pp. 1--19 (2007).

\bibitem{Dyer96}
J.~Dyer, {\it Organization Science\/} {\bf 7}, 649 (1996).

\bibitem{Zaheer98}
A.~Zaheer, B.~Mcevily, V.~Perrone, {\it Organization Science\/} {\bf 9}, 141
  (1998).

\bibitem{Uzzi97}
B.~Uzzi, {\it Administrative Science Quarterly\/} {\bf 42}, 35 (1997).

\end{thebibliography}

\bibliographystyle{Science}

\begin{scilastnote}
\item %
  We thank S. Wen, B. Li, and K. Nakajima for valuable discussions.
  This work was supported by JSPS KAKENHI Grant Number 24530506. Both authors
  have contributed equally to this work. 
\end{scilastnote}

\clearpage

\begin{sidewaystable}
\caption{{\bf Patent records and collaboration networks used in this
    paper}. The collaboration networks at the inventor and company levels are
  constructed from the Japan and U.S. patent records of several decades, with number of patents
  denoted by $N_\mm{P}$.  
  For each collaboration network we show 
  the number of nodes ($N$),
  edges ($L$), 
  mean degree ($\langle k \rangle = 2L/N$), 
  fraction of the largest connected component ($s_\mm{lc}$),
  fraction of isolated nodes ($n_\mm{0}$), 
  clustering coefficient ($C$) 
  and degree correlation ($r$).}
\centering
\vspace{0.1in}
\footnotesize
    \begin{tabular}{c|cc|rrrrrrr|rrrrrrr}
    \hline
    \hline
    & \multicolumn{2}{c|}{Patent record} & \multicolumn{7}{c|}{Inventor network} &
    \multicolumn{7}{c}{Company network} \\
    \hline
    & Duration & $N_\mm{P}$ & $N$ & $L$ & $\kmean$ & $s_\mm{lc}$ & $n_\mm{0}$ &  $C$ & $r$ 
                 & $N$ & $L$ & $\kmean$ & $s_\mm{lc}$ & $n_\mm{0}$ &  $C$ & $r$ \\
    \hline
Japan & 1994-2008 & 1,967,361 & 1,806,259 & 3,458,690 & 3.830 & 0.358 & 0.135 & 0.438 &  0.333
                  &    72,840 &    70,702 & 1.941 & 0.364 & 0.542 & 0.068 & -0.056 \\
\hline
USA   & 1963-1999 & 2,923,922 & 1,528,610 & 2,599,540 & 3.401 & 0.453 & 0.232 & 0.334 &  0.151
                  &   148,220 &    15,896 & 0.214 & 0.049 & 0.907 & 0.041 & -0.032 \\
   \hline
    \hline
    \end{tabular}\label{table:CN}
\end{sidewaystable}

\end{document}